\documentclass[conference]{IEEEtran}
\makeatletter
\def\ps@headings{%
\def\@oddhead{\mbox{}\scriptsize\rightmark \hfil \thepage}%
\def\@evenhead{\scriptsize\thepage \hfil \leftmark\mbox{}}%
\def\@oddfoot{}%
\def\@evenfoot{}}
\makeatother 

\IEEEoverridecommandlockouts
\usepackage{bbm}
\usepackage{amsfonts}
\usepackage[dvips]{graphicx}
\usepackage{times}
\usepackage{cite}
\usepackage{amsmath}
\usepackage{array}
\usepackage{amssymb}

\usepackage{stfloats}
\usepackage{slashbox}
\usepackage{graphicx}
\usepackage{footnote}
\usepackage{booktabs}
\usepackage{array}
\usepackage{algorithm}
\usepackage{subeqnarray}
\usepackage{cases}
\usepackage{threeparttable}
\usepackage{color}
\usepackage{epstopdf}
\usepackage{algpseudocode}
\usepackage{bm}
\usepackage{multirow}
\usepackage[labelformat=simple]{subcaption}
\usepackage{adjustbox}

\begin{document}\pagestyle{empty}

\abovedisplayskip = 2pt
\belowdisplayskip = 2pt

\title{Using Grouped Linear Prediction and Accelerated Reinforcement Learning for Online Content Caching}

\author{\authorblockN{Naifu Zhang, Kaibin Zheng and Meixia Tao}\thanks{This work is supported by the NSF of China under grant 61571299 and
61521062}
\authorblockA{Dept. of Electronic Engineering, Shanghai Jiao Tong University, Shanghai, China\\
Emails: \{arthaslery, kempin, mxtao\}@sjtu.edu.cn
}
\thanks{}
}

\maketitle
\vspace{-1.5cm}
\begin{abstract}
Proactive caching is an effective way to alleviate peak-hour traffic congestion by prefetching popular contents at the wireless network edge.
To maximize the caching efficiency requires the knowledge of content popularity profile, which however is often unavailable in advance.
In this paper, we first propose a new linear prediction model, named grouped linear model (GLM) to estimate the future content requests based on historical data.
Unlike many existing works that assumed the static content popularity profile, our model can adapt to the temporal variation of the content popularity in practical systems due to the arrival of new contents and dynamics of user preference.
Based on the predicted content requests, we then propose a reinforcement learning approach with model-free acceleration (RLMA) for online cache replacement by taking into account both the cache hits and replacement cost.
This approach accelerates the learning process in non-stationary environment by generating imaginary samples for Q-value updates.
Numerical results based on real-world traces show that the proposed prediction and learning based online caching policy outperform all considered existing schemes.
\end{abstract}
\section{Introduction}
Mobile data traffic has been experiencing an explosive growth in recent years, resulting in considerable traffic burden in both core networks and wireless access networks.
Proactive content caching at the edge of wireless networks has been shown to be an effective approach to alleviate the traffic burden, reduce content access latency, and improve user experience \cite{7537171,6871674}.
Due to the vast amount of contents available in multimedia platforms, not all of them can be pre-stored in local cache nodes.
Moreover, the content popularity is often unknown in advance.
It is therefore of great importance to predict the future content popularity and determine the cache placement strategy wisely.

One line of work on content popularity estimation assumes that the content popularity profile is unknown but time-invariant\cite{6871674,7387263,6883600,6874793,blasco2014content,6933489,7151068,7422747}.
In \cite{6871674,7387263} fixed global content popularity is estimated using a training set based on collaborative filtering and then exploited for caching decisions to maximize the average user request satisfaction ratio in small-cell networks.
In \cite{6883600,6874793,blasco2014content}, using multi-armed bandit algorithm, the cache node learns the content popularity distribution online by refreshing its cache placement and observing instantaneous demands for cached files.
In \cite{6933489}, diversity in content popularity across the user population is taken into account.
In \cite{7151068,7422747}, a transfer-learning approach is proposed to improve the popularity profile estimates by leveraging prior information obtained from a surrogate domain, such as social networks.
As mentioned, all these works \cite{6871674,7387263,6883600,6874793,blasco2014content,6933489,7151068,7422747} assumed that the content popularity profile is static and hence may not be applicable in practical systems where the content set or user preference are dynamic.

Another line of related work takes into account the temporal variation of the content set as well as user preference in cache placement.
The work \cite{7775114} learns the content popularity online using contextual multi-armed bandit optimization by taking into account the dependence of users¡¯ preference on their context.
The work \cite{yang2017dynamic} proposes a learning algorithm that estimates future content hit rate based on a linear prediction model.
This model incorporates content feature and location characteristics.
However, the users' context and content feature information exploited in \cite{7775114,yang2017dynamic} are often unavailable if the cache node is operated by mobile network operators, which in general can only observe local content requests.
The work \cite{7925848} integrates recommendation with local caching in wireless edge and proposes a reinforcement learning (RL) based algorithm to optimize the cache replacement policy.
A RL-based algorithm is also proposed in \cite{8241758} to find the optimal policy in an online fashion, enabling the cache node to track the space-time popularity dynamics.
Note that the action of caching in these existing RL-based algorithms is defined for each individual content and therefore the action space grows exponentially with the number of contents.

In this work, we propose a new prediction and learning approach for online content caching that can well suit the non-stationary environment without users' context or content feature information.
We assume that the content library is dynamic where new contents can arrive arbitrarily at each time slot.
We first propose a grouped linear model (GLM) to estimate the future content requests by treating the historical content requests as features where the linear coefficients are designed in a grouped manner according to the age of each available content.
To avoid over-fitting, instead of using regularization constraints, we impose linear constraints to parameters according to the correlation coefficient between the content requests.

Based on the predicted content requests by GLM, we propose a reinforcement learning approach with model-free acceleration (RLMA) to optimize the cache placement by taking both cache hit and the replacement cost into account.
We formulate the problem as a non-stationary Markov decision process (MDP), with the objective of maximizing the long-term reward.
Unlike the existing works on RL-based caching \cite{7925848,8241758}, we define the action as the number of contents, rather than the explicit set of contents, to be replaced, and hence the action space as well as state space are significantly smaller.
Instead of learning the environment model, we synthesize imaginary samples from historical information as inspired by the imagination rollouts method in \cite{gu2016continuous}.
We then update Q-values using imaginary samples, each of which has an adaptive learning rate.

We evaluate the performance of our prediction and learning algorithms using real-world traces.
Results show that the proposed RLMA algorithm in conjunction with the GLM provides higher long-term reward than existing policies.

\section{Problem Description}
We consider a content delivery network with one cache node of finite storage size.
We assume that new contents arrive in the system at arbitrary rates at discrete time slot $t=1,2,...$.
Let $\mathcal{N}_t$ denote a directory of newly arrival files at time slot $t$ with cardinality $|\mathcal{N}_t|=N_t$.
Let $\tau_f$ denote the release time slot (or birth time) of file $f$.
For each file $f\in\mathcal{N}_t$, its release time slot $\tau_f$ is $t$.
Let $\mathcal{F}_t$ denote a directory of total files at time slot $t$ with cardinality $|\mathcal{F}_t|=F_t$, which is empty at the beginning.
The content set is updated by $\mathcal{F}_t=\mathcal{F}_{t-1}\cup\mathcal{N}_t$.
Users can make random requests from $\mathcal{F}_t$ at time slot $t$, assuming that all files have the same size for convenience.

The cache node can store up to $M$ files from the file library.
Let $\mathbf{A}_t\in\mathcal{A}_t$ denote the $F_t\times1$ binary caching vector at time slot $t$ with $A_{t,f}=1$ indicating that the $f$-th file in $\mathcal{F}_t$ is cached at time slot $t$, and $A_{t,f}=0$ otherwise.
Here $\mathcal{A}_t=\{\mathbf{A}|\mathbf{A}\in\{0,1\}^{F_t},\mathbf{A^T 1}=M\}$ denotes the set of all feasible caching vectors satisfying the cache capacity constraint.
The replacement cost for refreshing caching files is considered.
In this paper, we define the replacement cost at time slot $t$ as the number of files that are cached at time slot $t$ but not at time slot $t-1$:
\begin{equation}
\begin{split}
C_t=\mathbf{A}_t^T[\mathbf{1}-\tilde{\mathbf{A}}_{t-1}],
\end{split}
\end{equation}
where $\tilde{\mathbf{A}}_{t-1}=[\mathbf{A}^T_{t-1},\mathbf{0}_{1\times N_t}]^T$.

Let $\mathbf{d}_t$ denote the content request vector at time slot $t$ of size $F_t\times 1$, where each element $d_{t,f}$ denotes the number of requests of file $f$ at time slot $t$.
If the requested content is cached, we consider that the file is hit.
We define the number of file hits at time slot $t$ as:
\begin{equation}
H_t=\mathbf{A}_t^T\mathbf{d}_t.
\end{equation}
If the cache node does not cache the requested files, it has to fetch the files from the content provider through the core network which increases the traffic load of the network and the access latency of the user.

Taking into account both the cache replacement cost and the cache hit, we define the cache utility at time slot $t$ as:
\begin{equation}
r_t=H_t-\lambda C_t,
\end{equation}
where $\lambda\geq 0$ is a weighting parameter of the replacement cost that accounts the affective factors such as network traffic, channel conditions, replacement speed, etc.
We aim to maximize the total cache utility over all time slots.
The optimization problem is given by,
\begin{equation}
\begin{split}
&\max\lim_{T\to\infty}\sum_{t=1}^T r_t\\
&\mbox{subject to:  }\mathbf{A}_t\in\mathcal{A}_t,t\in\{1,2,...,T\}.
\end{split}
\label{optimization}
\end{equation}
Since the instantaneous content requests $\mathbf{d}_t$ at each time slot cannot be revealed before making the caching decision $\mathbf{A}_t$, the above problem is intractable.

\begin{figure}[t]
\begin{centering}
\vspace{-0.2cm}
\includegraphics[scale=.30]{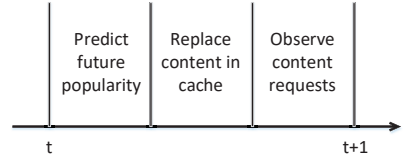}
\vspace{-0.1cm}
 \caption{\small{A schematic depicting the sequence of operations at time slot $t$}}\label{fig:slot}
\end{centering}
\vspace{-0.3cm}
\end{figure}

In this work, we propose a prediction and learning approach to solve (\ref{optimization}).
The sequence of system operations at each time slot $t$ is shown in Fig.~\ref{fig:slot}.
We first predict future content requests based on historical content requests.
Then we learn the content replacement policy to maximize the total cache utility and replace the cached contents accordingly.
At last we observe the content requests and compute the cache utility.

\section{Content Request Prediction Model}
A common way of predicting the request profile is to assume that content requests are generated under the independence reference model (IRM), where the request distribution for a given content is static and independent of all the past requests.
This IRM model is however not very realistic in practical systems.
Recently it is found in \cite{SNM} that the shot noise model (SNM) is more accurate than IRM since it can capture the temporal correlation of content requests. However, the SNM model assumes that the content requests are subject to exponential decay which in reality may not be the case.
In this work, we propose a new linear prediction model of the time-varying content requests by taking the dynamic aging of each file into account.

\subsection{Grouped Linear Model}
Define $b_{t,f}\triangleq t-\tau_f$, which indicates the age of content $f$ at time slot $t$.
Let $\mathbf{X}_{t,f}\in\mathbb{R}^{b_{t,f}}$ denote a $b_{t,f}$-dimensional feature vector of content $f$ observed since its release time, which is defined as:
\begin{equation}
\mathbf{X}_{t,f}=(d_{t-1,f},d_{t-2,f},...,d_{\tau_f,f})^T.
\end{equation}
The feature vector $\mathbf{X}_{t,f}$ contains the historical requests of content $f$ revealed from its release time slot $\tau_f$ to time slot $t-1$.
We predict content requests at time slot $t$ as a linear function of the feature vector.
The linear assumption is reasonable given the strong linear correlation between past and future popularities \cite{linear_assumption}.
Due to different dimensions of different feature vectors, the prediction of content requests at each time slot $t$ is classfied into different groups by the ages of the contents, $b_{t,f}$, with each having different combination coefficients.
More specifically, we predict the request of content $f$ at time slot $t$ as:
\begin{equation}
\hat{d}_{t,f}=\bm{\theta}_{b_{t,f}}\mathbf{X}_{t,f},
\label{predict}
\end{equation}
where $\bm{\theta}_{i}=(\theta_{i,1},\theta_{i,2},...,\theta_{i,i})\in\mathbb{R}^{i}$ is the unknown parameter vector for age $i=b_{t,f}$.

\subsection{Parameter Estimation}
Based on the grouped linear model in (\ref{predict}), we aim at finding the best choice of the parameter vectors $\Theta=\{\bm{\theta}_{1},\bm{\theta}_{2},...\}$ so as to achieve the optimal prediction accuracy.
The main challenge is to overcome the over-fitting problem.

Note that our proposed grouped linear model can be viewed as a variant of the Multivariate Linear Model (ML)\cite{ML} where regularization constraints (such as Ridge regression) are imposed to deal with the over-fitting problem.
However, the regularization constraints are for general purpose and do not utilize the specific meaning of the problem at hand.
In this work, content requests are both features and responses in the linear regression.
The parameter $\bm{\theta}_{i,j}$ can be viewed as a normalized correlation coefficient between the requests of contents with age $i$ at time slot $t$ and the requests of the same contents at time slot $t-j$, for $1\leq j\leq i$.
Since the content requests in the past encourage the content requests in the future, the correlation coefficient can be set to non-negative.
Moreover, the content requests during shorter time interval show stronger correlation.
Thus, we can further assume that $\bm{\theta}_{i,j}$ is monotonically non-increasing.
As such, we propose to estimate the optimal parameter vectors $\Theta$ by imposing linear constraints as follows:
\begin{align}
\mathcal{P}_1:~~\min_{\Theta}&\sum_{k=1}^{t}{\sum_{f\in\mathcal{F}_k}{(\hat{d}_{k,f}-d_{k,f})^2}}\nonumber\\
\mbox{s. t.}~~ &\bm{\theta}_{i,j}-\bm{\theta}_{i,j+1}\geq 0,~\forall j\in[1:i-1],~\forall\ i\in\mathbb{N}^+\nonumber\\
&\bm{\theta}_{i,i}\geq 0,~\forall i\in\mathbb{N}^+. \nonumber
\end{align}
\label{parameter}

The above linearly constrained optimization problem can be solved by gradient projection method.
Utilizing features with linearly constrained parameters endows our proposed GLM model with faster convergence and prevention of over-fitting.

\section{Learning-based Caching Policy Design}
In this section, we take content requests predicted by the GLM model as the actual content requests and propose a model-free adaptive reinforcement learning algorithm to optimize the caching policy in non-stationary environment.
\subsection{Q-Learning Formulation}
Before optimizing the caching strategy we first illustrate the tradeoff between the cache hits and the replacement cost.
On one hand, caching all the most popular $M$ files in time slot $t$ will maximize the hit rate, but it will also generate significant replacement cost due to the temporal dynamics of content requests.
On the other hand, if the caching vector remains unchanged with zero replacement cost, the hit rate will gradually decrease.

We resort to Q-learning to obtain the caching policy from problem (\ref{optimization}) for non-stationary Markov decision process (MDP) where the state and action are defined as follows:
\begin{itemize}
\item
{\bf State}:
The system state in each time slot is defined as
\begin{equation}
s_t=\mathbf{\hat A}_t^T[\mathbf{1}-\tilde{\mathbf{A}}_{t-1}].
\end{equation}
where $\mathbf{\hat A}_t$ denotes the anticipated caching vector that maximizes the cache hits at time slot $t$ and $\tilde{\mathbf{A}}_{t-1}$ is the zero-padded actual caching vector at time slot $t-1$ as defined earlier.
By intuition, $\mathbf{\hat A}_t$  should contain the $M$ most popular content at time slot $t$ based on the predicted content requests $\mathbf{\hat d}_t$.
The state $s_t$ represents the number of different contents between $\mathbf{\hat A}_t$ and $\mathbf{A}_{t-1}$, which limits the maximal number of contents to be replaced for maximizing the long-term system reward.
The overall state space is $\mathcal{S}=[0:M]$.
\item
{\bf Action}:
The action at each time slot $t$ is defined as the number of contents we should replace in the cache at time slot $t$, denoted as $a_t$.
Given the action $a_t$, since the replacement cost is fixed we should replace the least $a_t$ popular content in the caching vector $\mathbf{A}_{t-1}$ by the most popular $a_t$ contents not in the caching vector $\mathbf{A}_{t-1}$ in order to maximize the cache hits.
The action space at each time slot is $a_t\in [0:s_t]$.
\item
{\bf Reward}:
The reward at time slot $t$ is defined as the cache utility $r_t=H_t-\lambda C_t$.
\end{itemize}

We define the policy function $\pi :\mathcal{S}\rightarrow\mathcal{A}$, mapping any state $s\in\mathcal{S}$ to any action $a\in\{0,...,s\}$.
For the current state $s_t$, the caching vector is decided by action $a_t=\pi(s_t)$, indicating the number of contents replaced in the cache at time slot $t$.
Let $R_t(s,a)$ denote the average reward of executing action $a$ under state $s$ at time slot $t$.
Using $\pi$ as a complete decision policy, the optimal value of a state $s$ is written by
\begin{equation}
V_\pi(s_t)=\mathbb{E}\left [\sum_{\tau=t}^\infty{\gamma^{\tau-t} R_t(s_\tau,\pi(s_\tau))}\right ].
\end{equation}

The objective of adaptive Q-learning algorithm is to find the optimal policy $\pi^*$ such that the average reward of any state $s$ is maximized:
\begin{equation}
\pi^*=\arg \max_{\pi\in\Pi}{V_\pi(s)},\forall s\in\mathcal{S}
\label{policy}
\end{equation}
where $\Pi$ denotes the set of all feasible policy functions.

Bellman equation is a necessary condition for optimality of a policy in a sequential decision making problem.
Let $T_t(s,a,s')$ denote the transition probability that the current state $s$ goes to the next state $s'$ under action $a$ at time slot $t$.
In Markov decision processes, the expected reward for being in the state $s$ and following some fixed policy $\pi$ satisfies the Bellman equation:
\begin{equation}
V_{\pi }(s)=R_t(s,\pi(s))+\gamma\sum _{s'}T_t(s,\pi(s),s')V_{\pi }(s').
\end{equation}
Let $Q^*(s,a)$ be the expected discounted reinforcement of taking action $a$ in state $s$, then continuing by choosing actions optimally.
Note that $V^*(s)$ is the value of state $s$ assuming the best action is taken initially.
Hence, $V^*(s)=\max_a{Q^*(s,a)}$ and $Q^*(s,a)$ can be written recursively as
\begin{equation}
Q^*(s,a)=R_t(s,a)+\gamma\sum_{s'\in\mathcal{S}}{T_t(s,a,s')\max_{a'}{Q^*(s',a')}}.
\label{fullbackups}
\end{equation}

Note also that we have $\pi^*(s)=\arg\max_a{Q^*(s,a)}$ as an optimal policy.
Updates based on (\ref{fullbackups}) are known as full backups since they make use of information from all possible successor states.
We denote $(s,a,r,s',t)$ as an experienced sample summarizing a single transition in the environment.
This means that the agent in a state $s$, takes an action $a$, from which it receives a scalar reward $r$ from the environment at time slot $t$.
The reward $r$ is sampled with mean $R_t(s,a)$ and bounded variance.
The environment then goes to the next state $s'$ according to the distribution $T_t(s,a,s')$.
Given the $i$-th experienced sample $(s_i,a_i,r_i,s'_i,t_i)$, the Q-function at time slot $t$ is updated by applying a stochastic approximation as
\begin{equation}
Q_t(s_i,a_i)=(1-\alpha_{t})Q_t(s_i,a_i)+\alpha_{t}(r_i+\gamma\max_{a'}{Q_t(s_i',a')})
\label{modelfree}
\end{equation}
where $\alpha_{t}$ is the learning rate.
In stationary environment, the stochastic approximation algorithm \cite{stepsize} converges to the optimal value $Q^*$ when the learning rate $\alpha_{t}$ satisfies $\sum_{t=1}^\infty{\alpha_t}=\infty$ and $\sum_{t=1}^\infty{\alpha_t^2}<\infty$.
\subsection{Accelerated Reinforcement Learning}
In general, the content request profile is in a non-stationary environment due to the dynamic user preference and the release of new contents.
As such, the stochastic approximation algorithm (\ref{modelfree}) cannot converge to the optimal value $Q_t^*$.
When the learning rate of Q-learning tends to 0, the Q-function converges to a constant value but $Q_t^*$ is varying with time slot $t$ in non-stationary environment.

In contrast, constant learning rate is suitable for the dynamic $Q_t^*$, but there is still no guarantee of convergence mathematically because of insufficient experienced samples.
The sample complexity of model-free algorithms, particularly when producing merely one sample per time slot tends to limit the adaption.

In this work, we propose a novel approach to accelerate the convergence by synthesizing imaginary samples from historical information, and thereby increase the number of samples per slot.
Imagine that if we trace back at some point of time in the history and take an action different from what really happened in that time slot, then we can generate an imaginary reward and a new state in accordance with the known historical content requests.
By repeating this process we can get a sequence of imaginary samples.
However, the samples created in past slot do not follow the distribution of current reward and transition probability.
So instead of selecting constant learning rate, we propose an adaptive learning rate $\alpha_{t,i}=\alpha_0\beta_0^{t-t_i}$, where constant parameters $\alpha_0$ and $\beta_0$ satisfy $0<\alpha_0<1,0<\beta_0<1$.
The learning rate of updating sample decays exponentially further from the current slot.
The longer the sample is produced, the larger error of the average reward and transition probability the sample has.
In the extreme case when the sample is produced at the current slot, the learning rate achieves maximal value $\alpha_0$.
\subsection{Algorithm}
The pseudo code of our proposed reinforcement learning with model-free acceleration to re-adapt to the non-stationary environment is given in Algorithm \ref{Q-Learning}.
It starts with parameters to be initialized for the GLM model and reinforcement learning.
Replay buffer $\mathcal{B}$ stores experienced samples which are gained from real world, while imaginary replay buffer $\mathcal{B}_i$ stores imaginary samples.
In each time slot $t$, we traverse all the possible actions in current state $s_t$ to exploit the action $a_t$ with maximal value of Q-function.
Besides, after updating parameters of GLM model with observed content requests at current time slot $t$, we aggregate reward and store an experienced sample $(s_t,a_t,r_t,s_{t+1},t)$ in replay buffer $\mathcal{B}$.
We repeat synthesizing a sequence of imaginary samples in line 17-19 for $K$ times.
We first replace imaginary caching vector $\mathbf{A^*}_{t-\Delta t}$ with the real caching vector $\mathbf{A}_{t-\Delta t}$ in line 17, and then synthesize $\Delta t$ imaginary samples by repeating line 6-14.
Instead of taking action with maximal Q-value, we synthesize imaginary samples $(s_t,a_t,r_t,s_{t+1},t)$ by randomly taking actions allowing for maximizing exploration.
Adding the imaginary samples to the imaginary replay buffer $\mathcal{B}_i$ effectively augments the amount of samples available for Q-learning.
Then with samples in $\mathcal{B}$ and $\mathcal{B}_i$, we update Q-value and thus finish the learning process.

\begin{algorithm}[h]
\caption{Reinforcement Learning with Model-free Acceleration (RLMA)}
\label{Q-Learning}
\begin{algorithmic}[1]
\State Initialize normalized Q-value $Q(s,a)$ and caching vector $\mathbf{A}_0$;
\State Initialize replay buffer $\mathcal{B}\leftarrow\emptyset$ and step length $\Delta t\leftarrow 30$;
\State Initialize imaginary replay buffer $\mathcal{B}_i\leftarrow\emptyset$;
\State Initialize GLM parameters $\bm{\theta}\leftarrow 0$;
\For{$t=1,2,..,T$}
\State Observe state $s_t$;
\State Take action $a_t\leftarrow\arg \min_{a}{Q(s,a)}$;
\State Update the caching vector $\mathbf{A}_t$ by replacing $\mathbf{A}_{t-1}$ with $a_t$ files in $\mathbf{\hat A}_t$;
\State Content requests $\mathbf{d}_t$ are revealed;
\State Receive reward $r_t\leftarrow\lambda_r\mathbf{A}_t^T\mathbf{d}_t-\lambda_c\mathbf{A}_t^T[\mathbf{1}-\mathbf{A}_{t-1}]$;
\State Update GLM Model with $\mathbf{d}_t$ by $\mathcal{P}_1$;
\State Use GLM Model to predict $\mathbf{\hat d}_{t+1}$ by eq.(\ref{predict});
\State Compute $\mathbf{\hat A}_{t+1}$ by sorting $\mathbf{\hat d}_{t+1}$;
\State Compute $s_{t+1}$ by comparing $\mathbf{A}_t$ and $\mathbf{\hat A}_{t+1}$;
\State Store transition sample $(s_t,a_t,r_t,s_{t+1},t)$ in $\mathcal{B}$;
\For{$k=1,2,..,K$}
\State Set imaginary caching vector $\mathbf{A^*}_{t-\Delta t}\leftarrow\mathbf{A}_{t-\Delta t}$;
\State Repeat line 6-14 by replacing $\mathbf{A}$ with $\mathbf{A^*}$ for $\Delta t$ times;
\State Store all imaginary transition samples in $\mathcal{B}_i$;
\EndFor
\For{$(s_i,a_i,r_i,s_i',t_i)\in\mathcal{B}_i$}
\State Update $Q(s_i,a_i)$;
\EndFor
\EndFor
\end{algorithmic}
\end{algorithm}

\section{Simulation and Discussion}
In this section, we evaluate the performance of the proposed GLM model and RLMA algorithm by comparing them with several reference algorithms based on a real world dataset.
\subsection{Description of The Dataset}
We use data from MovieLens \cite{harper2016movielens} as our main dataset for the evaluation of our proposed algorithm.
MovieLens is an online recommender created by GroupLens Research in order to gather research data on personalized recommendations.
The MovieLens $224M$ DataSet \cite{ml-latest} contains $26024289$ ratings and $753170$ tag applications across 45843 movies.
These data were created by $270896$ users between January 09, 1995 and August 04, 2017.
Each entry of the dataset consists of an anonymous user ID, a movie ID, a rating made on a 5-star scale, and a timestamp.
We treat each movie rating by a user as a content request of a user connected to a wireless local cache node (see \cite{7775114,7524381} for a similar approach).
This assumption is reasonable since it is common that users post ratings on the movies right after watching them.

In our simulations, we only use the ratings of movies released in the last three years, since over half of the ratings were provided during this period.
Then, we divide three years' time into $1095$ slots of one day each $(T = 1095)$, assuming that the cache node updates its cache content on a daily basis.
Then, we assign the content requests to the slots according to their timestamps.

\subsection{Accuracy of Content Request Prediction}
We first evaluate the prediction accuracy of our proposed GLM method by considering the cache hits only without replacement cost.
We compare our method with the following benchmarks:
\begin{itemize}
\item
Least Recently Used (LRU):
The cache node records the recent arrive of all cached contents.
The least recently arrived one is replaced by the newly arrival content when the cache is full.
\item
Least Frequently Used with Dynamic Aging (LFUDA) \cite{806998}:
The cache node counts the the number of arrive of all contents.
The least frequently used one is replaced by the new content when the cache is full.
A dynamic aging policy is used to prevent previously popular content from polluting the cache.
\item
Optimal Caching:
The cache node caches the most popular content that achieves theoretically optimal performance with hindsight.
Note that this optimal caching only serves as the performance upper bound since it is not implementable in a real system because it needs future information.
\end{itemize}

\begin{figure}[t]
\begin{centering}
\vspace{-0.2cm}
\includegraphics[scale=.49]{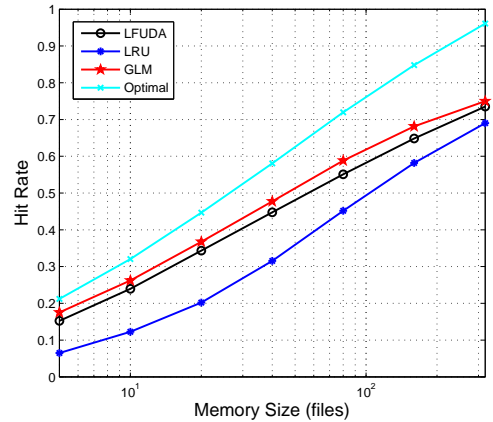}
\vspace{-0.1cm}
 \caption{\small{Overall cache hit ratio vs memory size}}\label{fig:Hitrate-M}
\end{centering}
\vspace{-0.3cm}
\end{figure}

Fig.~\ref{fig:Hitrate-M} plots the overall cache hit ratio with respect to cache size $M$ for a network with $T=1000$ time slots.
The overall cache hit ratio with cache size $M$ files is defined as $\sum_{t=1}^{T}{H_t}/\sum_{t=1}^{T}{\sum_{j=1}^{N}{d_{t,j}}}$.
It is seen that GLM outperforms both LFUDA and LRU.
In particular, the performance of GLM achieve $15.2\%$ improvement against LFUDA and $170.6\%$ improvement against LRU when the cache size is 5 files.
This is because LFUDA only considers static popularity profile which is different from the current popularity profile.
Although LRU takes the dynamics of popularity into account, the popular content may be replaced when unpopular content are requested.
The limitation of LRU becomes greater when the cache size is smaller.
Instead, GLM continuously learns the linear model not only considering the dynamic popularity but also avoiding the unstable replacement.

\begin{figure}[t]
\begin{centering}
\vspace{-0.2cm}
\includegraphics[scale=.49]{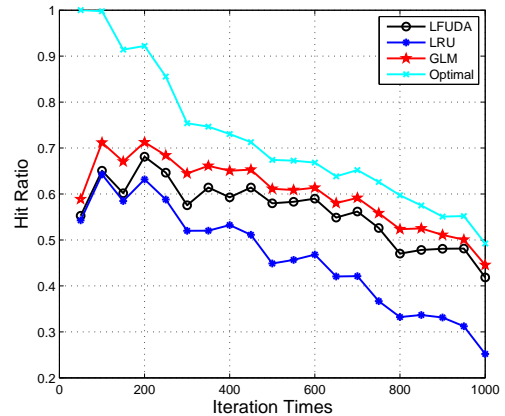}
\vspace{-0.1cm}
 \caption{\small{Average cache hit ratio vs iteration times}}\label{fig:Hitrate-t}
\end{centering}
\vspace{-0.3cm}
\end{figure}

Fig.~\ref{fig:Hitrate-t} plots the average cache hit ratio with respect to iteration times $t$ for a network with $M=80$ files to show how the caching
performance varies over time.
The instant cache hit ratio of time slot $t$ is define as $H_t/\sum_{j=1}^{N}{d_{t,j}}$.
Each point of a curve in the figure represents the average cache hit ratio within the time window between itself and the next point.
The average cache hit ratio achieved by these algorithm shown in this figure is decreasing.
This is due to the fact that the number of content requests for each time slot increases when the content library increases over time.
The performance of the GLM approaches the optimal policy over time, which indicates that GLM is good at identifying the most popular content instead of moderately popular content.

\subsection{Comparison of Caching Policies}
Now we consider the replacement cost and we evaluate the performance of our proposed RLMA algorithm.
Besides the benchmarks of LRU and LFUDA, we also compare the performance of RLMA algorithm with the following caching policies:
\begin{itemize}
\item
Most Popular:
The cache node caches the most popular content predicted by GLM algorithm.
The least popular one is replaced by the more popular content when the cache is full.

\item
Origin Q-Learning:
The origin Q-learning only updates the Q-function by the experienced samples instead of the imaginary samples.
\end{itemize}

\begin{figure}[t]
\begin{centering}
\vspace{-0.2cm}
\includegraphics[scale=.50]{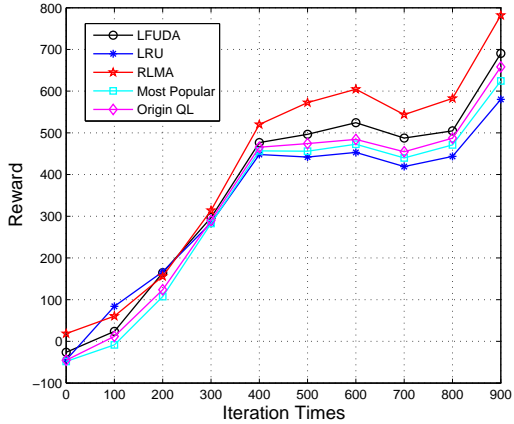}
\vspace{-0.1cm}
 \caption{\small{Average reward vs iteration times}}\label{fig:Replacement-Policy}
\end{centering}
\vspace{-0.3cm}
\end{figure}

Fig.~\ref{fig:Replacement-Policy} plots average reward with respect to iteration times $t$ for a network with cache size $M=1000$ files.
The cumulative reward achieved by RLMA is 15.6\%, 22.6\%, 28.8\% and 30.8\% higher than the ones achieved by LFUDA, Origin QL, Most Popular and LRU, respectively.
There are several notable points:
Origin QL only update its Q-function by the experienced samples, which cannot guarantee the convergence of the Q-value.
LFUDA performs badly at the first few hundred of slots because the popularity counted by LFUDA varies rigidly when the content library is small, which will cause a large replacement cost.
The performance of LRU is better than RLMA at first but drop to the worst after $400$ slots.
As the content library grows, the replacement cost of LRU increases because the uncached contents are more likely to be requested.

\section{Conclusion}
In this paper, we have considered a learning-based approach to cache contents in a single cache node.
The dynamic popularity profile for caching content is estimated using the grouped linear model and considers the historical content requests as features.
Linear constraints are added to the estimation to avoid over-fitting the training set.
By considering dynamic popularity profile and newly arrival contents, the content replacement is cast as a reinforcement learning for non-stationary environment task.
A reinforcement learning with model-free acceleration algorithm is developed for finding the optimal caching policy in an online fashion.
The imaginary samples achieve substantially improved sample complexity and enable the cache node to track and possibly adapt to the non-stationary environment.
Numerical results show that the proposed GLM Model outperforms LRU and LFUDA and the proposed RLMA algorithm provides higher long-term reward than other algorithms.
Further research may focus on multiple cache nodes, where each user can be served by one or multiple BSs.

\bibliographystyle{IEEEtran}
\bibliography{IEEEabrv,icc18ws}

\begin{thebibliography}{10}
\providecommand{\url}[1]{#1}
\csname url@samestyle\endcsname
\providecommand{\newblock}{\relax}
\providecommand{\bibinfo}[2]{#2}
\providecommand{\BIBentrySTDinterwordspacing}{\spaceskip=0pt\relax}
\providecommand{\BIBentryALTinterwordstretchfactor}{4}
\providecommand{\BIBentryALTinterwordspacing}{\spaceskip=\fontdimen2\font plus
\BIBentryALTinterwordstretchfactor\fontdimen3\font minus
  \fontdimen4\font\relax}
\providecommand{\BIBforeignlanguage}[2]{{%
\expandafter\ifx\csname l@#1\endcsname\relax
\typeout{** WARNING: IEEEtran.bst: No hyphenation pattern has been}%
\typeout{** loaded for the language `#1'. Using the pattern for}%
\typeout{** the default language instead.}%
\else
\language=\csname l@#1\endcsname
\fi
#2}}
\providecommand{\BIBdecl}{\relax}
\BIBdecl

\bibitem{7537171}
M.~Tao, W.~Yu, W.~Tan, and S.~Roy, ``Communications, caching, and computing for
  content-centric mobile networks: part 1 [guest editorial],'' \emph{IEEE
  Commun. Mag.}, vol.~54, no.~8, pp. 14--15, August 2016.

\bibitem{6871674}
E.~Bastug, M.~Bennis, and M.~Debbah, ``Living on the edge: The role of
  proactive caching in 5g wireless networks,'' \emph{IEEE Commun. Mag.},
  vol.~52, no.~8, pp. 82--89, Aug 2014.

\bibitem{7387263}
E.~Bastug, M.~Bennis, E.~Zeydan, M.~A. Kader, I.~A. Karatepe, A.~S. Er, and
  M.~Debbah, ``Big data meets telcos: A proactive caching perspective,''
  \emph{J. Commun. Networks}, vol.~17, no.~6, pp. 549--557, Dec 2015.

\bibitem{6883600}
P.~Blasco and D.~Gunduz, ``Learning-based optimization of cache content in a
  small cell base station,'' in \emph{Proc. IEEE ICC}, June 2014, pp.
  1897--1903.

\bibitem{6874793}
------, ``Multi-armed bandit optimization of cache content in wireless
  infostation networks,'' in \emph{Proc. IEEE ISIT}, June 2014, pp. 51--55.

\bibitem{blasco2014content}
------, ``Content-level selective offloading in heterogeneous networks:
  Multi-armed bandit optimization and regret bounds,'' \emph{arXiv preprint
  arXiv:1407.6154}, 2014.

\bibitem{6933489}
M.~S. ElBamby, M.~Bennis, W.~Saad, and M.~Latva-aho, ``Content-aware user
  clustering and caching in wireless small cell networks,'' in \emph{Proc.
  ISWCS}, Aug 2014, pp. 945--949.

\bibitem{7151068}
E.~Bastug, M.~Bennis, and M.~Debbah, ``A transfer learning approach for
  cache-enabled wireless networks,'' in \emph{Proc. WiOpt}, May 2015, pp.
  161--166.

\bibitem{7422747}
B.~N. Bharath, K.~G. Nagananda, and H.~V. Poor, ``A learning-based approach to
  caching in heterogenous small cell networks,'' \emph{IEEE Trans. Commun.},
  vol.~64, no.~4, pp. 1674--1686, April 2016.

\bibitem{7775114}
S.~M¨¹ller, O.~Atan, M.~van~der Schaar, and A.~Klein, ``Context-aware proactive
  content caching with service differentiation in wireless networks,''
  \emph{IEEE Trans. Wireless Commun.}, vol.~16, no.~2, pp. 1024--1036, Feb
  2017.

\bibitem{yang2017dynamic}
P.~Yang, N.~Zhang, S.~Zhang, L.~Yu, J.~Zhang, and X.~Shen, ``Dynamic mobile
  edge caching with location differentiation,'' \emph{arXiv preprint
  arXiv:1709.05377}, 2017.

\bibitem{7925848}
K.~Guo, C.~Yang, and T.~Liu, ``Caching in base station with recommendation via
  q-learning,'' in \emph{Proc. IEEE WCNC}, March 2017, pp. 1--6.

\bibitem{8241758}
A.~Sadeghi, F.~Sheikholeslami, and G.~B. Giannakis, ``Optimal and scalable
  caching for 5g using reinforcement learning of space-time popularities,''
  \emph{IEEE J. Sel. Topics Signal Process.}, vol.~PP, no.~99, pp. 1--1, 2017.

\bibitem{gu2016continuous}
S.~Gu, T.~Lillicrap, I.~Sutskever, and S.~Levine, ``Continuous deep q-learning
  with model-based acceleration,'' in \emph{Proc. ICML}, 2016, pp. 2829--2838.

\bibitem{SNM}
S.~Traverso, M.~Ahmed, M.~Garetto, P.~Giaccone, E.~Leonardi, and S.~Niccolini,
  ``Temporal locality in today's content caching: why it matters and how to
  model it,'' \emph{ACM SIGCOMM}, vol.~43, no.~5, pp. 5--12, 2013.

\bibitem{linear_assumption}
G.~Szabo and B.~A. Huberman, ``Predicting the popularity of online content,''
  \emph{Commun. ACM}, vol.~53, no.~8, pp. 80--88, 2010.

\bibitem{ML}
H.~Pinto, J.~M. Almeida, and M.~A. Gon{\c{c}}alves, ``Using early view patterns
  to predict the popularity of youtube videos,'' in \emph{Proc. ACM
  WSDM}.\hskip 1em plus 0.5em minus 0.4em\relax ACM, 2013, pp. 365--374.

\bibitem{stepsize}
V.~S. Borkar and S.~P. Meyn, ``The ode method for convergence of stochastic
  approximation and reinforcement learning,'' \emph{SIAM J. Control Optim.},
  vol.~38, no.~2, pp. 447--469, 2000.

\bibitem{harper2016movielens}
F.~M. Harper and J.~A. Konstan, ``The movielens datasets: History and
  context,'' \emph{ACM TIIS}, vol.~5, no.~4, p.~19, 2016.

\bibitem{ml-latest}
GroupLens, ``Movielens latest datasets,''
  \url{https://grouplens.org/datasets/movielens/latest/}.

\bibitem{7524381}
S.~Li, J.~Xu, M.~van~der Schaar, and W.~Li, ``Popularity-driven content
  caching,'' in \emph{Proc. IEEE INFOCOM}, April 2016, pp. 1--9.

\bibitem{806998}
J.~Dilley and M.~Arlitt, ``Improving proxy cache performance: analysis of three
  replacement policies,'' \emph{IEEE Internet Comput.}, vol.~3, no.~6, pp.
  44--50, Nov 1999.

\end{thebibliography}

\end{document}